 \documentstyle[12pt]{article}
\begin{document}
\baselineskip 5.5mm
 

 \newcommand{\be}[1]{\begin{equation}\label{#1}}
 \newcommand{\ee}{\end{equation}}
 \newcommand{\beqn}[1]{\begin{eqnarray}\label{#1}}
 \newcommand{\eeqn}{\end{eqnarray}}
\newcommand{\bd}{\begin{displaymath}}
\newcommand{\ed}{\end{displaymath}}
\newcommand{\mat}[4]{\left(\begin{array}{cc}{#1}&{#2}\\{#3}&{#4}\end{array}
\right)}
\newcommand{\matr}[9]{\left(\begin{array}{ccc}{#1}&{#2}&{#3}\\
{#4}&{#5}&{#6}\\{#7}&{#8}&{#9}\end{array}\right)}
\def\la{\lambda}
\def\al{\alpha}
\def\Ga{\Gamma}
\def\ga{\gamma}
 \newcommand{\eps}{\varepsilon}
\newcommand{\ov}{\overline}
\renewcommand{\to}{\rightarrow}
\def\mcirc{{\stackrel{o}{m}}}


\begin{flushright} 
hep-ph/9602326 ~~~~~~~~  INFN-FE 01/96 \\
January 1996 
\end{flushright} 
\vspace{1cm} 

\begin{center} 
{\bf ASTROPHYSICAL IMPLICATIONS OF THE MIRROR WORLD \\ [1mm]
WITH BROKEN MIRROR PARITY } \\ [0.3cm]

 Z.G. BEREZHIANI \footnote{
Talk given at the XIX int. conference on {\em Particle Physics and 
Astrophysics in the Standard Model and Beyond}, Bystra, Poland, 
19-26 September 1995 (to appear on Proceedings). 
Based on refs. \cite{BM,BDM} done in collaboration 
with A.D. Dolgov and R.N. Mohapatra. } \\ [0.3cm]

{\em INFN Sezione di Ferrara, I-44100 Ferrara, Italy, and \\ 
Institute of Physics, Georgian Academy
of Sciences, Tbilisi, Georgia  } \\ [0.5cm]

\end{center}

\begin{abstract} 
We discuss the physics of the mirror (shadow) world which is 
completely analogous to the visible one except that its `weak' scale 
is larger by one or two orders of magnitude than the weak scale in 
the standard model. The mirror neutrinos 
can mix the ordinary ones through the Planck scale induced higher 
order operators, which can help to reconcile the present 
neutrino puzzles that are the solar and
atmospheric neutrino deficits, the recent LSND anomaly 
and the need in the $\sim$ eV mass neutrino as the hot dark matter. 
In particular, the oscillation of $\nu_e$ into its mirror partner 
$\nu'_e$ emerges with parameters naturally in the MSW range. 
The nucleosynthesis constraint on the extra light particle species
can be fulfilled by assuming the asymmetric postinflationary 
reheating between the usual and mirror worlds. 
One implication of our proposal is that 
 bulk of the dark matter in the universe may 
be a warm dark matter consisting of the keV range mirror neutrinos
rather than the conventional cold dark matter, while the mirror 
baryons can also contribute as dissipative dark matter. 
Implications of the mirror Machos for microlensing experiments  
are  also discussed.
\end{abstract}


\vspace{0.4cm}
\begin{center}
{\bf 1. Neutrino Puzzles} 
\end{center}

Direct measurements show no evidence for any of the neutrinos to be 
massive. However, there have been indirect ``positive'' signals 
for neutrino masses and mixing accumulated during the past years. 
These are: 

A. {\it Solar neutrino problem} (SNP). The  solar $\nu_e$ 
deficit indicated by the solar neutrino experiments \cite{SNP} 
cannot be explained by nuclear or astrophysical reasons 
(see \cite{Gianni} and references therein). 
The most popular and natural solution is provided by the 
MSW oscillation of $\nu_e$ into another neutrino 
$\nu_x$ in solar medium \cite{MSW}.
It requires the oscillation parameters in the range 
$\delta m^2_{ex}\sim 10^{-5}~\mbox{eV}^2$ and  
$\sin^2 2\theta_{ex}\sim 10^{-3}-10^{-2}$.
Another possible solution is related to 
the long wavelength ``just-so'' oscillation 
from Sun to Earth \cite{just-so}, which requires a parameter range 
$\delta m^2_{ex}\sim 10^{-10}~\mbox{eV}^2$ and 
$\sin^2 2\theta_{ex}\sim 1$. 

B. {\it Atmospheric neutrino problem} (ANP). 
There is an evidence for a significant depletion of the atmospheric 
$\nu_\mu$ flux by almost a factor of 2 \cite{ANP}. 
This points to $\nu_\mu-\nu_x$ 
oscillation, with $\delta m^2_{\mu x}\sim 10^{-2}~\mbox{eV}^2$ 
and $\sin^2 2\theta_{\mu x}\sim 1$.

C. {\it Dark matter problem} (DMP).  
The COBE measurements of the cosmic microwave
background anisotropy suggests that cosmological dark
matter consists of two components, cold dark matter (CDM) being
a dominant component 
and hot dark matter (HDM) being a smaller admixture \cite{HDM,HDM2}.  
The latter role can be naturally played by neutrinos with mass of
few eV's. As for the CDM, there are several candidates, e.g. 
the lightest supersymmetric particle (LSP) or the axion condensate. 
However, it is of certain interest to think of it as also 
consisting of some heavier (keV range) neutrinos 
with correspondingly small concentration, 
so called warm dark matter (WDM) \cite{WDM}.

D. {\it LSND result:}
Direct evidence of $\bar{\nu}_\mu-\bar{\nu}_e$ oscillation
from the recent Los Alamos experiment \cite{LSND}, with
$\delta m^2_{e\mu}\geq 0.3~\mbox{eV}^2$ and
$\sin^2 2\theta_{e\mu}= 10^{-3}-10^{-2}$.

If all these hints (or at least first three of them) 
will indeed be confirmed in future experiments, then three standard  
neutrinos  $\nu_e,\nu_{\mu}$ and $\nu_{\tau}$ would not suffice 
for their explanation. Since existence of the fourth active neutrino 
is excluded by the LEP measurements of the invisible decay width of 
$Z$-boson, one has to introduce an extra light sterile 
neutrino $\nu_s$. 
It was shown \cite{calmoh} that only one possible texture is compatible 
with all the above mentioned data, which requires that 
$m_{\nu_{e,s}}\ll m_{\nu_{\mu,\tau}}$. In this case the SNP can be  
explained by the $\nu_e-\nu_s$ oscillation and the ANP 
by the $\nu_{\mu}-\nu_{\tau}$ oscillation. In addition, 
the $\nu_{\mu}$ and $\nu_{\tau}$ with mass $\simeq 2.4$ eV 
provide the cosmological HDM and can  also explain the LSND result. 
%

One can question,  from where the sterile neutrino comes from  
and how it can be so light when it is allowed to have a large mass 
by the gauge symmetry of the standard model. 
We suggest \cite{BM} that the sterile neutrino is in fact
a neutrino of a shadow world which is the mirror duplicate
of our visible world, but its `electroweak' scale $v'$ is by a 
factor of $\zeta\sim 30$ larger than the standard electroweak scale $v$. 
Thus, the mirror neutrinos $\nu'_{e,\mu,\tau}$ should be light 
by the same reason as the usual ones $\nu_{e,\mu,\tau}$: 
their mass terms are suppressed by the accidental B-L symmetry 
resulting from the gauge symmetry and field content 
of the theory.\footnote{ 
For some other proposals for light $\nu_s$ invoking 
extra global symmetries, see \cite{SV,Chun}. }  

This framework can provide a plausible solution to 
all present neutrino puzzles. We suppose that  
the dominant entries ($\sim$ eV) in the neutrino mass matrix 
have origin in some intermediate scale physics which respects an  
approximate global ZKM-type lepton number 
$\bar L=L_e+L_\mu-L_\tau$ conservation in both sectors. 
This fixes the neutrino mass texture
needed for reconciling the HDM requirement, 
the ANP and the LSND oscillation,   
while the $\nu_e$ and $\nu'_e$ states are left massless.   
The masses and mixing of the latter and thereby the 
oscillation $\nu_e-\nu'_e$ then can be induced by the 
Planck scale effects \cite{BEG,ABS}, which 
for $\zeta\sim 30$ properly reproduce the parameter range 
needed for the MSW solution to the SNP. 
The question arises, what is a possible role played by two 
other mirror neutrinos $\nu'_{\mu,\tau}$. 
In the framework presented below they have masses in
the keV range and constitute the WDM of the universe.


The concept of the hidden mirror world has been considered in 
several earlier papers \cite{mirror}. A key difference of our 
approach \cite{BM,BDM} 
from the earlier ones is that we consider a case of the 
spontaneously broken mirror parity between two worlds, 
so that the weak scales $v$ and $v'$ are different. 
This can allow also to reconcile the Big Bang 
nucleosynthesis (BBN) constraints on the effective number of the 
light neutrinos \cite{NB}. 
The mirror photons and neutrinos could apriori give a large 
contribution considerably exceeding $N_\nu=3$. Therefore, 
their abundances at the BBN epoch must be 
appropriately reduced. To achieve this goal, we assume an asymmetric 
inflationary reheating between the two universes, which can naturally 
occur once the mirror parity is spontaneously broken. In particular, if 
the mirror universe reheats to a lower temperature than our universe, 
then BBN constraint can be satisfied.  
We also address some cosmological implications of the mirror particles.

\vspace{2mm}
\begin{center}
{\bf 2.  Within the Visible World }
\end{center}

Apparently, three known neutrinos are not enough to explain all 
the present neutrino puzzles \cite{calmoh}. 
The key difficulty is related to the SNP, while the 
other puzzles can be easily reconciled. One can assume that  
the neutrino mass matrix in flavour basis $\nu_{e,\mu,\tau}$ 
has a texture obeying the approximate 
$\bar L = L_e + L_\mu - L_\tau$ conservation: 
\be{texture}
\hat{m}_\nu = \matr{0}{0}{m\sin\theta_{e\mu}}
{0}{0}{m\cos\theta_{e\mu}}
{m\sin\theta_{e\mu}}{m\cos\theta_{e\mu}}{\eps m} 
\ee
with say $m\simeq 2.4$ eV and $\eps\simeq 10^{-3}$. 
Then it has one massless eigenstate $\nu_1\simeq \nu_e$ and 
two almost degenerate eigenstates  
$\nu_{2,3}\simeq \frac{1}{\sqrt{2}}(\nu_\mu\pm \nu_\tau)$, 
with masses $m_{2,3}\approx m(1\pm \frac{\eps}{2})$. 
The latter will play a role of the HDM \cite{HDM2}, while 
the ANP can be explained by the oscillation $\nu_\mu - \nu_\tau$ 
with $\sin 2\theta_{\mu\tau}\approx 1$ 
and $\delta m^2_{\mu\tau} \approx 2\eps m^2\simeq 10^{-2}$ eV$^2$,  
and  the oscillation $\bar{\nu}_\mu - \bar{\nu}_e$ with 
$\delta m^2_{e\mu}=m^2 \simeq 6\,\mbox{eV}^2$ can
explain the LSND result, if $\sin\theta_{e\mu}\simeq 0.02$.
Thus, only the SNP remains unresolved.

One can built a simple seesaw model that could 
naturally implement the above structure. Let us introduce 
beyond the left-handed lepton doublets $l_{e,\mu,\tau}$ and 
the right-handed charged leptons $e,\mu,\tau$ of the standard model, 
also two right-handed neutrinos $N$ and $S$ (subscripts $L$ and $R$ 
are omitted). We prescribe the global lepton number 
$\bar L=1$ to states $l_{e,\mu}$, $e,\mu$ and $N$, and 
$\bar L=-1$ to states $l_{\tau}$, $\tau$ and $S$. 
Let us also introduce a gauge singlet scalar $\Phi$ with $\bar L=-2$.  
Then the relevant Lagrangian has a form 
\beqn{Lagr} 
(g_e \ov{l}_e \phi e + g_\mu \ov{l}_\mu \phi \mu + 
g_\tau \ov{l}_\tau \phi \tau) + 
(g_1 \ov{l}_e \tilde{\phi} N + g_2 \ov{l}_\mu \tilde{\phi} N + 
g_3 \ov{l}_\tau \tilde{\phi} S) \nonumber \\ 
~~~~~~ + M N C S + g_N\Phi NCN + g_S \Phi^{\ast} SCS 
\, + \, {\rm h.c.} 
\eeqn  
where $\phi$ is a standard Higgs doublet with a VEV $v=174$ GeV  
($\tilde{\phi}=i\tau_2 \phi^\ast$), and $C$ is the charge conjugation 
matrix. We assume that the scalar $\Phi$ developes a nonzero VEV 
$\langle \Phi \rangle=V \ll M$, which spontaneously breaks the 
$\bar L$-invariance and gives rise to majoron in the particle spectrum. 
We also assume that the Yukawa constants $g_{N,S}\sim 1$ while 
$g_{1,2,3}$ have the same pattern as $g_{e,\mu,\tau}$, so that 
the Dirac mass terms  of neutrinos emerge with approximately the 
same magnitudes as masses of the charged leptons: 
$m^D_{1,2,3}\sim m_{e,\mu,\tau}$.  
Clearly, this model provides a texture resembling (\ref{texture}). 
One neutrino eigenstate $\nu_1$ (an admixture of $\nu_e$ and $\nu_\mu$ 
with angle $\theta_{e\mu}= m^D_1/m^D_2 \sim 10^{-2}$) 
is left massless while two other eigenstates $\nu_{2,3}$ get masses  
$m\simeq m^D_2m^D_3/M$ through the seesaw mixing to the 
heavy states $N,S$, with small splitting  
$\eps \sim (m^D_3/m^D_2)(V/M)$.    

Then $m$ in the eV range corresponds to $M\sim 10^8$ GeV, while 
$\eps \sim 10^{-3}$ demands that $V \sim 10^4$ GeV. 
Interestingly, the latter value satisfies the bound on the 
spontaneous lepton number breaking scale which arises if the 
the Planck scale effects on the majoron are taken into account 
\cite{ABMS}. This is important, since the Planck scale effects 
will play a crucial role in our further considerations. 

\vspace{2mm}
\begin{center}
{\bf 3. Introducing the Mirror World } 
\end{center}

Having in mind the $E_8\times E_8'$ string theory, 
one can imagine that in the field-theoretical limit it reduces 
to a gauge theory given 
by the product of two identical groups $G\times G'$,  
where $G=SU(3)\times  SU(2)\times U(1)$ stands for the 
standard model describing particles of the visible world: 
quarks and leptons $q_i, ~u^c_i, ~d^c_i; ~l_i, ~e^c_i$ and 
the Higgs doublet $\phi$,  
and $G'=[SU(3)\times SU(2)\times U(1)]'$ is its mirror 
counterpart with analogous particle content: 
$q'_i, ~u'^c_i, ~d'^c_i; ~l'_i, ~e'^c_i$  and $\phi'$ 
($i=1,2,3$ is a family index).
Let us suppose also that there exists a discrete mirror parity 
$P(G\leftrightarrow G')$ interchanging 
all particles in corresponding representations of $G$ and $G'$. 
It implies that all coupling constants 
(gauge, Yukawa, Higgs) have the same pattern in both sectors.  
Let us also assume that 
there is some mechanism that spontaneously breaks $P$ parity 
at lower energies and thus allows the weak interaction scales 
$\langle \phi \rangle =v$ and $\langle \phi' \rangle =v'$ 
to be different; below we assume that $v'/v \sim 30$. 
Thus the fermion mass and mixing pattern in the mirror world 
is completely analogous to that of the visible world, but 
all fermion masses are scaled up by the factor $\zeta=v'/v$. 
The $W',Z'$ boson and mirror Higgs  masses are also scaled up by 
factor $\zeta$,  
while photons and gluons remain massless in both systems. 

We suppose that two worlds communicate only through gravity 
and possibly also via some superheavy gauge singlet matter. 
It is also essential that at higher energies 
the $SU(3)\times SU(2)\times U(1)$ factors 
in both sectors should be embedded into some simple gauge groups: 
otherwise, kinetic terms of the two $U(1)$ gauge fields can mix which
would induce arbitrary electric charges to the particles \cite{mirror}.  
E.g. one can consider (SUSY) GUT   
like $SU(5)\times SU(5)'$ without mixed representations, 
which then breaks down to $G\times G'$ at higher energies. 
We also assume that the mirror parity $P$ is not broken at the 
GUT scales and its breaking is essentially related to the electroweak 
symmetry breaking scales $v$ and $v'$. 

Concerning the strong interactions, it is clear that 
a big difference between the weak scales $v'$ and $v$ 
will not cause as big difference between the confinement scales 
in two worlds. As far as $P$ parity is valid at higher scales, 
the strong coupling constant will evolve down in energies 
with same value in both sectors until it reaches threshold of 
the mirror-top ($t'$) mass.
Below it $\alpha'_{s}$ will have a different slope than $\alpha_{s}$.
It is then very easy to calculate the value of the scale $\Lambda'$ 
at which $\alpha'_{s}$ becomes large. 
This value will of course depend on $\zeta$. 
By taking $\Lambda=200$ MeV for the ordinary QCD, then 
for $\zeta\sim 30$ we find $\Lambda' \simeq 280$ MeV. 
On the other hand, for $\zeta\sim 30$ the masses of `light' mirror 
quarks $m'_{u,d}=\zeta m_{u,d}$ 
do not exceed the value of $\Lambda'$ and thus 
they should develope condensates $\langle \bar{q}' q' \rangle$ 
as their visible partners do. 
So, the mirror $\pi$-mesons should have masses rather 
close to that of our $K$-mesons. 
Also the mirror nucleons are not much heavier than the usual 
proton and neutron, $m'_{p,n}\sim 1.5 m_{p,n}$.  
However, the mirror light quarks have $\sim 200$ MeV masses 
and thus we expect the mass difference between the 
mirror neutron and proton of 100 MeV or so, while the mirror 
electron mass is $m'_e=\zeta m_e \sim 15$ MeV. 
Then unlike in our world, in the mirror
world all bound neutrons will be unstable against $\beta$ decay 
and the mirror hydrogen will be the only stable nucleus.

As for neutrinos, they can acquire nonzero masses only via 
operators bilinear in the Higgs fields and cutoff by some large scale 
$M$. As far as the $P$ parity breaking is a lower 
energy phenomenon, it should be respected by these operators. 
For example, one can directly extend the model of previous section, 
by introducing along with the heavy neutral states $N,S$ 
also their mirror partners $N',S'$, with the same mass 
$M\sim 10^{8}$ GeV. 
Then the neutrino mass operators in two sectors are:  
\be{LRnu}
\frac{h_{ij}}{M} (l_i\phi)C(l_j\phi) +
\frac{h_{ij}}{M} (l'_i\phi')C(l'_j\phi') + {\rm h.c.} 
\ee 
with constants $h_{ij}$ 
obeying the approximate $L_e+L_\mu-L_\tau$ symmetry. 
In doing so, the $\nu_{2,3}$ states get almost degenerate 
masses $m\sim$ few eV and thus can constitute HDM, and the ANP and 
LSND problems are also solved. 
Then for $\zeta\sim 30$, mass of their mirror partners $\nu'_{2,3}$, 
$m'=\zeta^2 m$, is in the keV range  
and thus the latter could consitute the WDM of the universe. 

The $\nu_e$ and $\nu'_e$ states are left massless. 
However, they can get masses from the Planck scale effects which 
explicitly violate the global lepton number, and 
can also induce the neutrino mixing between two sectors \cite{ABS}. 
The relevant operators are:  
\be{Planck}
\frac{\alpha_{ij}}{M_{Pl}} (l_{i}\phi)C(l_{j}\phi) +
\frac{\alpha_{ij}}{M_{Pl}} (l'_{i}\phi')C(l'_{j} \phi')
+ \frac{\beta_{ij}}{M_{Pl}} (l_i\phi)C(l'_{j} \phi') + {\rm h.c.}
\ee
with constants $\alpha,\beta\sim 1$.
Then $\nu_e$ and $\nu'_e$ acquire masses respectively 
 $\sim \mcirc$ and  $\sim \zeta^2 \mcirc\ $, 
and their mixing term is  $\sim \zeta \mcirc $, 
where $\mcirc\ =v^2/M_{Pl}= 3\cdot 10^{-6}~ \mbox{eV}$.
Hence, parameters of the oscillation $\nu_e-\nu'_e$ are in the range: 
\be{alter}
\delta m^2\sim \left(\frac{\zeta}{30}\right)^4 
\times 8\cdot 10^{-6}\,\mbox{eV}^2, ~~~~~
\sin^2 2\theta\sim  
\left(\frac{30}{\zeta}\right)^2 \times 5\cdot 10^{-3} 
\ee
which for $\zeta\sim 30$ perfectly fits 
the  small mixing angle MSW solution to the SNP \cite{MSW}.
More generally, by taking into account the solar model uncertainties 
\cite{Gianni,MSW}, as well as the possible order of magnitude spread 
in constants $\alpha,\beta$, the relevant range for $\zeta$ can be 
extended to $\zeta\sim 10-100$ \cite{BM}.  Alternatively, 
for $\zeta\sim 1$ we get $\delta m^2 \sim 10^{-10}\,\mbox{eV}^2$ 
and $\sin^2 2\theta \sim 1$, which corresponds to the 
`just-so' solution \cite{just-so}.

\vspace{2mm}
\begin{center}
{\bf 4. Spontaneous Parity Breaking and Asymmetric Inflation } 
\end{center}

The simplest possibility to spontaneously break the mirror parity 
$P(G\leftrightarrow G')$  is to introduce a $P$-odd real scalar 
$\eta$ with VEV $\langle \eta \rangle = \mu$ at some intermediate 
scale \cite{CMP}. Then the Higgs potential has a form:  
\beqn{eta} 
{\cal V}(\eta)\, + \,
m^2 (|\phi|^2 + |\phi'|^2)\, + \, 
\lambda (|\phi|^4 + |\phi'|^4) \nonumber \\
+ \, g\mu \eta (|\phi|^2 - |\phi'|^2)\, + \,
h\eta^2 (|\phi|^2 + |\phi'|^2) 
\eeqn 
so that after the non-zero VEV of $\eta$ emerges, the effective 
mass terms of $\phi$ and $\phi'$ become different and their VEVs 
$v$ and $v'$ will be different as well. As for the gauge and Yukawa 
coupling constants, they will not be affected and thus will 
maintain the same pattern in both worlds. 
Hence, the particle spectrum in the mirror world 
will have the same shape as that of the visible one but 
scaled up by the factor $\zeta=v'/v$.  

On the other hand, the $P$-parity breaking can be related 
to the possibility of asymmetric postinflationary ``reheating'' 
between two worlds.\footnote{The idea of using inflation to provide 
a temperature difference between ordinary matter and mirror or other 
forms of hidden matter was first discussed in ref. \cite{KST}. } 
In fact, it is natural to assume that the field $\eta$ itself plays 
a role of the inflaton \cite{BM,BDM}, 
provided that potential ${\cal V}(\eta)$ is sufficiently flat
and  satisfies all necessary `inflationary' conditions \cite{Linde}. 
As far as $P$-parity is broken, then it should be violated 
also in the inflaton couplings to visible and mirror Higgses:  
e.g. constants of trilinear couplings of the inflaton to $\phi$ and 
$\phi'$  become respectively $(h\pm g)\mu$. 
Then the inflaton will decay into the visible and mirror particles 
with different rates, so that the two thermal bathes can be 
established having different temperatures $T_R$ and $T'_R$. 
One has also to assume that already at the 
reheating stage two worlds are decoupled from each other.
In particular, this means that couplings like 
$a|\phi|^2 |\phi'|^2$, in principle allowed by symmetry, 
should be strongly suppressed: $a< 10^{-7}$. 
In this way, the initial cosmological abundance of the mirror 
particles can be smaller than that of their 
visible partners.
For more detailed discussion of asymmetric reheating 
and possible realistic supersymmetric models see ref. \cite{BDM}. 

Let us discuss now constraints on the difference of the reheating 
temperatures $T_R$ and $T'_R$. 
The most serious bound emerges 
from the BBN constraint on the effective number of the light 
particle species \cite{NB}. 
As far as the two worlds are decoupled already at the inflationary 
reheating epoch, during the universe expansion they evolve with 
separately conserved entropy\footnote{In fact, this 
applies if the expansion goes adiabatically in both sectors 
and only the second order phase transitions occur. 
At the presence of the first order phase transitions 
in both sectors this relation would change due to additional 
entropy production \cite{BDM}. }.  
Then the $T_R,T'_R$ are related to the temperatures $T,T'$ 
respectively of the usual and mirror photons 
at the BBN era as 
\be{T_R} 
\frac{T'_R}{T_R} =  
\left(\frac{2+5.25x^3}{10.75}\right)^{\frac{1}{3}}\frac{T'}{T} \,, 
~~~~~~~~~~ x =\frac{T'_\nu}{T'}
\ee 
where $T'_\nu$ is a temperature of the mirror neutrinos.

In standard cosmology effective number of the light degrees 
of freedom at the BBN era is $g_\ast=10.75$ as it is contributed 
by photons, electrons and three neutrino species $\nu_{e,\mu,\tau}$,  
in a remarkable agreement with 
the observed abundances of light elements \cite{NB}.  In our case, 
the mirror photons $\gamma'$ and neutrinos $\nu'_{e,\mu,\tau}$
would also contribute the effective number of the light particles, as 
\be{BBN}
\Delta g_\ast = 1.75\Delta N_\nu = 
(2 + 5.25 x^4) \left(\frac{T'}{T}\right)^4 
\ee 
The value of $x$ is determined by the temperature $T'_D$ at which 
$\nu'$ decouple from the mirror plasma.  It approximately scales as 
$T'_D \sim \zeta^{4/3} T_D$,  where $T_D=2-3$ MeV is the decoupling 
temperature of the usual neutrinos. 
For smaller values of $\zeta$, when  $T'_D<\Lambda'$,  
$\nu'$ decouple after the mirror QCD phase transition, so that 
the mirror electrons $e'$ contribute 
the heating of $\gamma'$ after the $\nu'$ freezing out and we 
arrive to the standard result $x=(4/11)^{1/3}=0.71$. 
For sufficiently large $\zeta$,  
$T'_D$ can be larger than both the QCD scale $\Lambda'$ and the 
light mirror quark masses $m'_{u,d}=\zeta m_{u,d}$.  
Then $u',d'$ and the mirror gluons will also contribute 
and we obtain $x=(4/53)^{1/3}=0.36$. 
Thus, for $x$ in the interval $0.71-0.36$ from eqs. (\ref{BBN}) 
and (\ref{T_R})  we obtain: 
\be{T'} 
\frac{T'}{T} < (0.85-0.96) (\Delta N_\nu)^{1/4} ~~~~ \Longrightarrow 
~~~~ \frac{T'_R}{T_R} < (0.60-0.57) (\Delta N_\nu)^{1/4}
\ee
Therefore, by taking the conservative bound $\Delta N_\nu < 1$  
or very strong limit $\Delta N_\nu <0.1$, we obtain that 
reheating temperature in the mirror sector 
has to be about 2 or 3 times smaller than that of the visible world.

A somewhat stronger bound can be derived from the overclosure 
constraint of the universe. Since in our model almost degenerate
$\nu_\mu$ and $\nu_\tau$ have masses $m\sim$ few eV, they form 
the HDM of the universe with $\Omega_{\nu}=2m/(94 h^2 ~\mbox{eV})$. 
Then their mirror partners $\nu'_{\mu,\tau}$ being $\zeta^2$ times 
heavier would contribute the cosmological energy density 
as  $\Omega'_{\nu} =r\zeta^2 \Omega_\nu$, 
where $r=n'_\nu/n_\nu$ stands for the present abundance of mirror 
neutrinos relative to ordinary ones: $r=(xT'/T)^3$, 
and $h$ is the Hubble constant in units 100 Km s$^{-1}$ Mpc$^{-1}$.   
Then by taking rather conservatively that 
$\Omega_\nu + \Omega'_\nu < 1$ (bearing in mind that other particles 
like LSP or mirror baryons could also contribute the present 
energy density), we obtain 
\be{WDM}
\frac{T'}{T} < \frac{1}{x \zeta^{2/3}} 
(\Omega_\nu^{-1} -1)^{1/3} = 
0.29 \left(\frac{0.36}{x}\right)\left(\frac{30}{\zeta}\right)^{2/3} 
(\Omega_\nu^{-1} - 1)^{1/3} 
\ee
Therefore, if $\Omega_\nu\simeq 0.2$ \cite{HDM}, 
then for $\zeta\approx 30$ we get that 
$T'/T< 0.46$ and $T'_R/T_R< 0.27$, which limits 
are comparable to that of eqs. (\ref{T'}) for $\Delta N_\nu =0.05$!

\vspace{2mm}
\begin{center}
{\bf 5. Cosmological Implications of the Mirror Particles } 
\end{center}

Thus, the concept of the mirror universe 
suggests a natural possibility for solving the DMP with 
dark matter components entirely consisting of neutrinos. 
Indeed, in the most interesting case the electroweak scale $v'$
in the mirror sector should be by factor $\zeta\sim 30$ larger than 
the standard electroweak scale $v=174$ GeV, while the
reheating temperature of the mirror universe
should be 3-4 times smaller than that of the visible one.
Hence, if the usual neutrinos with $m \sim $ few eV form the 
HDM component, then bulk of the dark 
matter can be the WDM component consisting of their  
mirror partners with $m'=\zeta^2 m \sim $ few keV.  
Clearly, the latter could form the halo dark matter even in dwarf 
spheroidal galaxies where the Tremaine-Gunn limit 
is most stringent ($m'_\nu> 0.3-0.5$ keV).

Implications of the WDM for the shape of the large scale 
structure are rather similar \cite{WDM} to that of the 
currently popular CDM made upon the heavy ($m\sim 100$ GeV) 
particles or axion condensate. 
%
However, more detailed observational data on the distribution of 
matter in the universe may make it possible to
discriminate between warm and cold dark matter.
Moreover, dark matter consisting of sterile neutrinos invalidates
direct searches of the CDM candidates via superconducting detectors
or axion haloscopes. High energy neutrino fluxes from
the Sun and from the Galactic center
which are expected from the annihilation of LSP's if they dominate
in the universe, will also be absent.
In supersymmetric versions of our scheme, however, CDM  as well 
could exist in the form of the LSP.

An interesting question is what is the amount and form of
the mirror baryonic dark matter in the universe.
Most likely, baryogenesis in the mirror
universe proceeds through the same mechanism as in the visible one
and we may expect that the baryon
asymmetries in both worlds should be nearly the same.
Since mirror nucleons are not much heavier than the usual ones, 
their fraction in the present energy density,
$\Omega_{B'}$, would be about the same as $\Omega_B$,
that is around a few percent.

Let us discuss now cosmological evolution of mirror baryonic matter. 
Since the binding energy of the mirror hydrogen atom
is thirty times larger than that of the ordinary hydrogen, its
recombination occurs much before the usual recombination era.
Hence, the evolution of density fluctuations in the mirror matter
would be more efficient than in the visible one.
(From the viewpoint of the visible observer
mirror baryons behave as a dissipative dark matter.)
As a result, one can expect
that the distribution of mirror baryons in galactic discs
should be more clumped towards the center. It is noteworthy that
mirror dark matter may show antibiasing behaviour ($b<1$) which
is considered unphysical for normal dark matter.
Recent data on the dark matter distribution in dwarf spiral galaxies 
obtained at smaller distances from the center and with a better 
resolution, do not agree with the assumption of purely collisionless 
dark matter and may indicate the existence of dissipative dark 
matter \cite{dwarf}.  

On the other hand, 
since mirror hydrogen is the only stable nucleus in the mirror world,
nuclear burning could not be ignited and
luminous (in terms of $\gamma'$) mirror stars cannot be formed.
Therefore, nothing can prevent the sufficiently big protostars to 
collapse and in dense galactic cores a noticeable fraction of 
mirror baryons should form the black holes.
Recent observational data indeed suggest a presence of giant black
holes with masses $\sim 10^{6-7} M_\odot$ in galactic centers.
In addition, easier formation of mirror black holes may explain the
early origin of quasars. 

The remaining fraction of the mirror baryons
could fragment into smaller objects like  white dwarves
(or possibly neutron stars) which can maintain stability
due to the pressure of degenerate fermions.
For the mirror stars consisting entirely of hydrogen,
the Chandrasekhar limit is
$M'_{\rm Ch}= 5.75 (m_p/m'_p)^2 M_\odot \simeq 3 M_\odot $. 
For smaller mirror objects the evaporation limit
should be $2-3$ orders of magnitude smaller than for the
visible ones because the Bohr radius of the mirror hydrogen is
30 times smaller than that of the usual one.

These mirror objects, being dark for the normal observer,
could be observed as Machos in the gravitational microlensing
experiments (for a review, see e.g. ref. \cite{Ansari}).
In principle they can be distinguished from the Machos of the 
visible world. The latter presumably consist of the dim compact
objects (brown dwarves) too light to burn hydrogen, with
masses ranging from the evaporation limit
$(\sim 10^{-7} M_\odot)$ to the ignition limit
$(\sim 10^{-1} M_\odot)$ \cite{DeRuj}. 
As for the mirror Machos, their mass spectrum can extend 
from the smaller evaporation limit $\sim 10^{-9} M_\odot$
up to the Chandrasekhar limit $\sim 3M_\odot$.
The present data on the microlensing events
are too poor to allow any conclusion on the presence of such
heavy (or light) objects. An unambiguous determination of the Macho
mass for each event is impossible, and only the most probable mass
can be obtained, depending of the spatial and velocity distribution
of Machos. The optical depth or the fraction of the sky covered
by the Einstein disks of Machos,  is nearly
independent of their mass: the Einstein disk surface
is proportional to $M$, while the number of deflectors for a given
total mass decreases as $M^{-1}$.
However, larger event statistics will allow to find
the Macho mass distribution with a better precision.

As noted earlier, the distribution of mirror baryonic matter
in galaxies should be more shifted towards their centers as compared
to the visible matter. Thus one can expect that
mirror stars in our Galaxy would significantly contribute to the
microlensing events towards the galactic bulge, while their weight 
in the microlensing events in halo should be smaller than that
of usual Machos. Interestingly, the event rates in the galactic
bulge observed by OGLE and MACHO experiments are about twice
larger than the expected value deduced from the low mass star population
in the Galactic disk \cite{OGLE}.
Barring accidental conspiracies like a
presence of bar (elongated dense stellar distribution along the
line of sight), this can be explained by the contribution of mirror
stars, which could naturally increase the optical depth towards
the galaxy bulge by factor 2 or so.



\vspace{5mm}
\noindent
{\bf Acknowledgments}
\vspace{3mm} 

I thank Sasha Dolgov and Rabi Mohapatra for collaboration on this  
subject. I also wish to thank  Jan Sladkowski, Marek Zralek and other 
organizers  for the most pleasant hospitality 
during the beautiful school in Bystra, 
and R. Ansari, J. Bahcall, D. Caldwell, M. Krawczyk, H. Mayer, 
G. Vitiello and other participants of the school for interesting 
and intensive discussions around these issues.

\end{document}